\documentclass[10pt, conference, compsocconf]{IEEEtran}
\usepackage{url,comment,cite}
\usepackage{amsmath,amssymb,amsfonts,bbm}
\usepackage{graphicx}
\usepackage{epsfig,epsf,psfrag,textcomp,tabularx}
\usepackage[usenames,dvipsnames]{xcolor}
\usepackage{subfigure}
\usepackage{float}
\usepackage{stfloats,mathtools}
\usepackage[algoruled,medskip,dontprintsemicolon,linesnumbered,Algorithm]{algorithm}
\usepackage[noend]{algpseudocode}

\newcommand{\Cc}{\mathcal{C}}
\newcommand{\Ac}{\mathcal{A}}
\newcommand{\Mc}{\mathcal{M}}
\newcommand{\Nc}{\mathcal{N}}
\newcommand{\Ec}{\mathcal{E}}

\numberwithin{equation}{section}

\begin{document}

\title{Real-Time Scheduling\\
for Content Broadcasting in LTE}

\author{\IEEEauthorblockN{Francesco Malandrino}
\IEEEauthorblockA{
Politecnico di Torino\\
Torino, Italy\\
Email: malandrino@tlc.polito.it
}
\and
\IEEEauthorblockN{Claudio Casetti}
\IEEEauthorblockA{
Politecnico di Torino\\
Torino, Italy\\
Email: casetti@tlc.polito.it
}
\and
\IEEEauthorblockN{Carla-Fabiana Chiasserini}
\IEEEauthorblockA{
Politecnico di Torino\\
Torino, Italy\\
Email: chiasserini@tlc.polito.it
}
\and
\IEEEauthorblockN{Siyuan Zhou}
\IEEEauthorblockA{
Politecnico di Torino\\
Torino, Italy\\
Email: siyuan.zhou@polito.it
}
}

\maketitle
\thispagestyle{empty}
\pagestyle{empty}

\begin{abstract}
Broadcasting capabilities are one of the most promising features of upcoming LTE-Advanced networks.
However, the task of scheduling  broadcasting sessions is far from trivial, since it affects
the available resources of several contiguous cells as well as the
amount of resources that can be devoted to unicast traffic. In this paper, we
present a compact, convenient model for broadcasting in LTE, as well as a set of efficient algorithms
to define broadcasting areas and 
to actually perform content scheduling. 
We study the performance of our algorithms in a realistic scenario, deriving interesting
insights on the possible trade-offs between effectiveness and computational efficiency.
\end{abstract}

\section{Introduction
\label{sec:intro}
}

LTE and LTE-Advanced networks have been conceived and designed for the purpose of facing an ever-increasing
demand for capacity. Indeed, smartphones and tablets are now full-fledged entertainment stations, capable of
displaying high-quality multimedia content -- and their owners seem to love that. The days when the expression
``mobile multimedia'' referred to playing hiccup-plagued cat videos from YouTube are long gone; users demand
low-latency multiplayer gaming, real-time video uploading and, increasingly, high-definition streaming.

Streaming is an especially challenging use case, in that it requires both high speed and low latency. Too many
users playing the same content can choke even a high-capacity network such as LTE. Even high-capacity networks
such as LTE can have trouble supporting too many users playing the same stream.
Several approaches have been proposed to tackle this challenge \cite{noi-infocom14,hetnets}. 
%Some~\cite{coop-stream} involve cooperative
%downloading: a user downloads the content from LTE, and then relays it to others through
%device-to-device (D2D) technologies such as Bluetooth. In a similar vein, more recent
%proposals~\cite{noi-infocom14} involve network-controlled, in-band D2D transfers. Furthermore, supporting
%streaming is one of the foremost use cases motivating the so-called heterogeneous networks~\cite{hetnets},
%where macro- and pico-cells coexist within the same LTE network.
In addition to general-purpose techniques such as heterogeneous networking, there is a proposed feature of LTE
that targets exactly the issue of real-time streaming -- {\em broadcasting}. The intuition behind it is simple:
operators can decide to devote a part of their spectrum resources to broadcasting high-demand content. Users
requesting the content will be served without further increasing the network load, just as it happens with
DVB television. Small-scale experiments involving broadcasting to mobile devices through LTE
have been successfully carried out~\cite{superbowl}, and mobile operators are planning to employ this technology
for massively popular sport events such as the Super Bowl~\cite{superbowl}. More specifically, operators have to decide:
\begin{itemize}
\item {\em whether} to broadcast a certain content at all;
\item {\em when} to broadcast it, accounting for its current (and future) popularity;
\item {\em where} to broadcast it, as popularity is typically location-specific.
\end{itemize}
Such decisions must be taken in a clever way -- broadcasting a content that is not sufficiently popular implies
wasting precious spectrum resources, which could be used to serve ordinary, i.e., unicast, traffic. 
Of equal importance, and perhaps of a more challenging nature, decisions must be swift.

%\begin{figure}
%\centering
%\includegraphics[width=0.9\columnwidth]{img/logical_arc}
%\caption{Broadcasting in LTE: logical architecture. Solid lines indicate
%content; dashed lines indicate control and signaling.
%\label{fig:logical-arc}}
%\end{figure}

%\begin{figure}[b!]
%\centering
%\subfigure[]{
%\includegraphics[width=0.3\textwidth]{img/network-example}
%}
%%\subfigure[]{
%%\includegraphics[width=0.3\textwidth]{img/subframe}
%%}
%\caption{
%Broadcasting in LTE: example network.
%\label{fig:lte-time}
%}
%\end{figure}

Taking swift decisions concerning LTE broadcasting is difficult for several reasons. The most obvious is that
the elements to account for -- potential content, associated demand, unicast traffic -- change rapidly over time.
Furthermore, the decisions that have to be taken are complex: deciding to broadcast a content in a cell has
far-reaching consequences in terms of interference on the neighboring ones. Finally, there are technology-
and standard-related constraints to honor, e.g., concerning the maximum amount of resources that can be
devoted to broadcasting.

The solutions that have appeared in the literature so far have aimed
at solving the problem of where massively-popular content should be
broadcasted 
\cite{Rong,Alexiou}. 
In addition to network configuration, significant attention has been devoted to
scheduling and resource allocation \cite{SVC}. Indeed, in LTE broadcasting, UEs can send
feedback about their perceived quality of service, and such information can be
leveraged to adjust the scheduling over time. 
Finally, the white paper in~\cite{EricssonWP} describes an early implementation of LTE broadcasting,
and its ability to improve the network capacity and performance.

In this paper, we chart the path for the broadcasting of non-massively-popular content on LTE networks. 
Our contribution is twofold.
First, we present a model for broadcasting in LTE networks. Simple and compact as it is, our model can
capture all the decisions that have to been taken, their consequences and implications, and the constraints
they are subject to. After discussing the impracticality of solving such a model to the optimum, we
make our second contribution: a family of scheduling algorithms that are:
\begin{itemize}
\item effective, in that their output is close to the optimum;
\item efficient, in that such an output is computed in a short time;
\item informed, in that  they account for the consequence of scheduling decisions on unicast traffic, as well as for the existing constraints.
\end{itemize}
Finally, we assess the effectiveness of our algorithms in a large-scale, realistic scenario.

The remainder of this paper is organized as follows. 
We describe how broadcasting is implemented in LTE standards in 
Section~\ref{sec:lte-bcast}. We present
our model in Section~\ref{sec:model}, and our algorithms in
Section~\ref{sec:algos}. We study their effectiveness in
Section~\ref{sec:results}. Lastly, Section~\ref{sec:conclusions}
concludes the paper.

\section{Broadcasting in LTE
\label{sec:lte-bcast}
}

3GPP has introduced
MBMS (Multimedia Broadcast and Multicast Service)
as a point-to-multipoint way to broadcast and multicast data to mobile
users, in release R6~\cite{MBMS-R6}. %Due to inter-symbol interference (ISI) issues,
In LTE system, MBMS has evolved into enhanced MBMS, 
with the introduction of Single-Frequency Networks (SFNs),
in release R9~\cite{36.300V9}.
%Further improvements have been introduced in R10~\cite{36.300V10}.
%,
%most notably a counting procedure to keep track of how many users
%are interested in each content. Service continuity for
%users moving across cells has been introduced in R11~\cite{36.300V11}.
%Broadcasting support has been introduced in LTE to provide efficient
%point-to-multipoint services, where a single network entity (e.g., an eNodeB)
%sends data to multiple UEs.
%LTE broadcasting takes place in-band, i.e., on licensed LTE frequencies, and
%is directed by the infrastructure. 
%More specifically, as shown in
%Fig.~\ref{fig:logical-arc}, each eNB is served by a
%Multi-cell/multicast coordination entity (MCE), in charge of admission
%control and radio resources allocation. Actual packets are sent through
%IP multicast by the MBMS gateway.
%The most important feature of LTE broadcasting is that it is
%implemented through SFNs.
All eNBs belonging to the same SFN
transmit the same information (the same bits) on the same carrier
frequencies (licensed bands), in a
synchronized fashion. This prevents interference  within the same SFN.

Each SFN can span multiple {\em contiguous} cells; the set of cells belonging to the
same SFN is called {\em MBSFN area} (Multimedia Broadcast over SFN). The maximum number
of allowed MBSFN is 256 per geographical region.

Time multiplexing is another important aspect. A first constraint is that
at most 6 out of 10 subframes can be used for broadcasting.
Furthermore, UEs cannot be expected to receive data
from multiple MBSFN areas at the same time. However, UEs can belong to multiple areas;
it follows that the schedules of overlapping MBSFN areas cannot overlap.

%Fig.~\ref{fig:lte-time} presents an example of such a time division. The cell in the
%middle of the topology belongs to two areas; this means that the combined length of
%the scheduling in the two areas cannot exceed six subframes.

\section{System model
\label{sec:model}
}

Our model focuses on a single time frame, during which it is
reasonable to assume that aspects such as time variations
in content popularity and user mobility do not vary.
Without loss of generality, 
we assume that in each broadcasting area, it is broadcast exactly one 
piece of content. This 
means that, if we need to broadcast more than one content in the same
cells, multiple areas will be created.

\subsection{Building blocks}

\subsubsection{Cells and areas}

There are three main components of our system: cells, broadcast areas, and content.

Cells~$c\in\Cc$ are standard LTE cells. 
We call $\Ec\subseteq\Cc^2$ the set that contains all pairs of neighboring cells;
thus $(c_1,c_2)\in\Ec$ if $c_1$ and $c_2$ are neighbors.
To simplify the notation, we write the set of neighbors of cell~$c$ as $\Nc^c=\{c^{\prime}\in\Cc\colon(c,c^{\prime})\in\Ec\}$.
%\begin{equation}
%\nonumber
%\Nc^c=\{c^{\prime}\in\Cc\colon(c,c^{\prime})\in\Ec\}.
%\end{equation}
Notice that~$c$ is considered a neighbor of itself, i.e.,~$c\in \Nc^c$. Also, in each cell~$c$ there are a total
of~$U^c$ users.

Areas~$a\in\Ac$ are the broadcast areas that we create and  correspond to the MBSNF areas
in LTE. To comply with LTE limitations, it should be:~$|\Ac|\leq
256$. Clearly, the size of any area cannot exceed the total number of
cells in the region, i.e., $|a|\leq |\Cc|$. 

\subsubsection{Content and popularity}

We denote by ~$m\in\Mc$\ the content items we may decide to broadcast, e.g., live events.
For each cell~$c$ and content item~$m$, we know the popularity~$\pi^c_m$, i.e., the number of users
in cell $c$ interested in content item~$m$ at the current time.

\subsubsection{Resources}

Spectrum resources correspond to LTE resource blocks (RBs), and represent the usage we are making
of the LTE spectrum.
For each content item~$m$, we know the amount~$\rho^c_m$ of resources needed to 
broadcast item~$m$
in cell~$c$. Such an amount depends on both the content (e.g., the
video resolution) and the cell, e.g., propagation conditions
experienced by its users.

The number of existing resources is~$R$, out of which at most~$r\leq R$ are available for broadcasting.
This allows us to represent, e.g., the 6/10 limit discussed in Section~\ref{sec:lte-bcast}.

\subsection{Decision variables}

We have two main decisions to take: which cells belong to each area, and which content is broadcasted
in each area. To this end, we define two decision variables:
\begin{itemize}
\item a binary variable~$\alpha_a^c\in\{0,1\}$, expressing whether cell~$c\in\Cc$ belongs to area~$a\in\Ac$;
\item a discrete variable~$\mu_a\in\Mc$, expressing which content is broadcasted in area~$a\in\Ac$.
\end{itemize}

Furthermore, we define an auxiliary variable~$x_a$, expressing the amount of resources we use for
broadcasting within area~$a\in\Ac$. As discussed next, we need this variable in order to account for
technology and standard constraints.

\subsection{Constraints}

The first constraint we need to impose concerns the minimum amount of resources~$\rho^c_m$. If cell~$c$
belongs to area~$a$, i.e.,~$\alpha^c_a=1$, then area~$a$ must use enough resources to properly stream
its content~$\mu_a$ to cell~$c$:
\begin{equation}
\label{eq:constr-mu}
x_a\geq\rho^c_{\mu_a},\forall a\in\Ac,c\in\Cc\colon\alpha^c_a=1.
\end{equation}

Next, we account for 
the total amount~$r$ of resources that can be devoted to broadcasting. For each cell~$c\in\Cc$, the sum of the~$x$-values of the cells
it belongs to cannot exceed~$r$:
\begin{equation}
\label{eq:constr-R}
\sum_{a\in\Ac\colon\alpha^c_a=1}x_a\leq r,\forall c\in\Cc.
\end{equation}
Note that constraint~(\ref{eq:constr-R}) also poses a soft limit to the number of areas a cell can belong to.

Finally, we have to deal with interference. The most conservative approach would impose that if a resource
is used by area~$a$ in cell~$c$, then it should not be used by any other area overlapping either $c$
or any cell neighboring~$c$. A softer constraint is given by:
\begin{equation}
\label{eq:constr-R-ex}
\sum_{a\in\bigcup_{c^{\prime}\in \Nc^c}\left\{a\in\Ac\colon\alpha^{c^{\prime}}_a=1\right\}}x_a\leq r,\forall c\in\Cc.
\end{equation}
As complex as it looks, constraint~(\ref{eq:constr-R-ex}) has a simple meaning: for each cell~$c$, the areas to
which~$c$ or any of its neighbors belong should have enough resources
available so that a disjoint set can be scheduled. Recall that~$c$ is
included in~$\Nc^c$ by definition.

\subsection{Performance metric and objective}

Intuitively, our goal is to set the~$\alpha$- and~$\mu$-variables so as to maximize the system performance.
However, the definition of ``system performance'' is rather vague, and deserves a deeper discussion.

Let us focus on a cell~$c$, with~$U^c$ users within it. Also, let~$\Ac^c=\{a\in\Ac\colon \alpha^c_a=1\}$ be the
set of areas~$c$ belongs to, and~$\Mc^c=\{\mu_a,\forall a\in \Ac^c\}$ the set of content broadcasted therein.

We can identify three distinct groups of users:
\begin{itemize}
\item users that are served through broadcasting;
\item users that would like broadcast-able content, but are not served by broadcasting;
\item users that want to download unicast content.
\end{itemize}
For each group of users, we can compute a satisfaction metric.

Users that are served through broadcasting have satisfaction~$1$. The number of such users is given by $\sum_{m\in \Mc^c}\pi^c_m$.
%\begin{equation}
%\nonumber
%\sum_{m\in \Mc^c}\pi^c_m.
%\end{equation}
The remaining users are served through unicast. The pool of resources that can be assigned to them is
given by the total amount~$R$, minus the ones used by the area(s) that cell~$c$ belongs to, minus the ones
interfered by neighboring cells: $R-\sum_{a\in\Ac\colon\sum_{c^{\prime}\in \Nc^c}\alpha^{c^{\prime}}_a= 1}x_a$.
%\begin{equation}
%\nonumber
%R-\sum_{a\in\Ac\colon\sum_{c^{\prime}\in \Nc^c}\alpha^{c^{\prime}}_a= 1}x_a.
%\end{equation}

With these resources, we have to serve the users that request content in~$\Mc\setminus \Mc^c$, i.e.,
not transmitted through broadcast.
The total amount of resources needed by these users is $\sum_{m\in\Mc\setminus \Mc^c}\pi^c_m\rho^c_m$.
%\begin{equation}
%\nonumber
%\sum_{m\in\Mc\setminus \Mc^c}\pi^c_m\rho^c_m.
%\end{equation}
%We assume that these users, as well as those interested in unicast
%content, will be requesting a GBR (Guaranteed BitRate) bearer: either
%they will be given the resources they need, hence have
%satisfaction~$1$, or get nothing, and have satisfaction~$0$. 
By denoting the number of unicast users and their resource request in
cell $c$ by
$\pi^c_u$ and $\rho^c_u$, respectively, the 
average number of satisfied users is given by:
\begin{equation}
\nonumber
\frac{
R-\sum_{a\in\Ac\colon\sum_{c^{\prime}\in \Nc^c}\alpha^{c^{\prime}}_a= 1}x_a.
}{
\sum_{m\in\Mc\setminus \Mc^c}\pi^c_m\rho^c_m + \pi^c_u \rho^c_u
}.
\end{equation}

Combining the above expressions, we can define the following
performance metric:
\begin{multline}
\label{eq:performance}
V^c=\sum_{m\in \Mc^c}\pi^c_m %+\\
+\frac{
R-\sum_{a\in\Ac\colon\sum_{c^{\prime}\in \Nc^c}\alpha^{c^{\prime}}_a= 1}x_a.
}{
\sum_{m\in\Mc\setminus \Mc^c}\pi^c_m\rho^c_m + \pi^c_u \rho^c_u
}.
\end{multline}
Equation~(\ref{eq:performance}) takes values between~$0$ and the
number of users on our topology, and it represents the average total number
of satisfied users.

%\subsection{Complexity issues}
%
%In our formulation, the problem belongs to the Mixed-Integer Linear Programming (MILP) 
%category. Such problems
%are NP-hard, i.e., their computational complexity grows exponentially with the problem size (e.g.,
%the number of variables)~\cite{linopt}. It follows that it is not feasible to solve it for 
%realistically-sized instances.
%
%Our next step is therefore to seek an alternate solution to the problem, trading some limited
%amount of accuracy for a massive improvement in terms of scalability.

\section{Our approach
\label{sec:algos}
}
As mentioned earlier, we have two main tasks to perform:
\begin{itemize}
\item assigning the cells to the areas;
\item deciding the content to broadcast in each area.
\end{itemize}
Jointly tackling these tasks would require solving a MILP problem that, as discussed above,
is intractable for realistic instances of the problem.

Therefore, we resort to a {\em divide-et-impera} approach \cite{MILCOM}, and decouple the two tasks.
Specifically, we present a simple, efficient way to select the
content~$\mu_a$ to broadcast in an area~$a$
given the cells belonging to it, i.e., the~$\alpha^c_a$ values. We leverage
such an assignment technique and reduce our scheduling problem to assigning
cells to the area, i.e., setting the~$\alpha$-values.

\subsection{Selecting the content}

Here, we assume we already know the cell-to-area assignment, i.e., the~$\alpha^c_a$-values
for all areas~$a\in\Ac$ and cells~$c\in\Cc$.
Our task is to determine the content~$\mu_a\in\Mc$ to broadcast in each area.

We proceed in a straightforward way, as summarized in Algorithm~\ref{alg:decide-content}.
We begin by ranking areas by the number of users interested in
broadcastable content that they  include (line~\ref{line:cont-sort}).
Then, for each area starting from the biggest one, 
we identify the viable content, i.e., content that can be broadcasted to that area
without violating constraint~(\ref{eq:constr-R-ex}) (line~\ref{line:cont-viable}).
Finally (line~\ref{line:cont-argmax}), we select the viable content that maximizes
the overall performance, as defined in~(\ref{eq:performance}).

Notice that it is possible (line~\ref{line:cont-continue}) that the set of viable
content is empty, i.e., no content can be broadcasted in the area. A typical reason
for this is that all the 6/10 subframes available for broadcasting are occupied by other
areas overlapping (or neighboring) with the current one. In this case, we simply
proceed with the next area.

\begin{algorithm}
\begin{algorithmic}[1]
\Require{$\Cc,\Ac,\{\alpha^c_a$\}}
\State{\texttt{sort}~$a\in\Ac$~by~no. of users interested in broadcasting} \label{line:cont-sort}
\ForAll{$a\in\Ac$}
\State{{viable\_content\_set}$\gets\{m\in\Mc\colon$(\ref{eq:constr-R-ex})~holds$\}$} \label{line:cont-viable}
\If{{viable\_content\_set}$\equiv\emptyset$}
\State{{\bf continue}} \label{line:cont-continue}
\EndIf
\State{$\mu_a\gets\arg\max_{\mbox{\rm viable\_content\_set}} V^c$} \label{line:cont-argmax}
\EndFor
\end{algorithmic}
\caption{Assigning content to areas
\label{alg:decide-content}}
\end{algorithm}

In Algorithm~\ref{alg:decide-content} we follow a greedy approach,
i.e., we never reconsider decisions once they are taken. This means that we have no formal
optimality guarantee. However, starting from the areas with the
highest number of users interested in broadcastable content, guarantees that any conflicts
are solved in such a way that the largest number of users is satisfied.

Finally, we remark that, while solving the MILP formulation to the optimum in
small-scale scenarios, we noticed that the selection of content~$\mu_a$ has a
smaller 
impact on the system performance
than the cell-to-area assignment. It is thus preferable to employ a straightforward
approach for selecting content, as we did, and a more sophisticated solution for the
area formation.

\subsection{Forming the areas}

Different cells have, in general, different demand for different content. Intuitively,
the most straightforward action one could take is forming as many areas as there are cells, with
each area comprising one cell. Two factors concur in rendering such a straightforward
solution undesirable and, in the general case, infeasible: the maximum number of areas that can be created, e.g., 256,
and the inter-area interference.
%\begin{itemize}
%\item the maximum number of areas that can be created, e.g., 256;
%\item inter-area interference.
%\end{itemize}

The first aspect is clear: there is a hard limit on the number of
areas we can form.
Inter-area interference is a bit more complex. As mentioned in Section~\ref{sec:lte-bcast},
areas are implemented as single-frequency networks; therefore, there is no interference
between cells belonging to the same area. Neighboring areas, instead, are subject to
interference; we model this through constraint~(\ref{eq:constr-R-ex}) in Section~\ref{sec:model}.

It follows that if we have two neighboring cells with similar (albeit not identical)
content popularity values~$\pi^c_m$, it is often better to put them in the same area (and
serve only the content item that is popular in both cells) than having two separate areas
whose schedules are tailored to each cell.

%\begin{table}[h!]
%\caption{Example: content demand and resource requirements
%\label{tab:example}}
%\centering
%\renewcommand{\arraystretch}{1.2}
%\begin{tabularx}{1\columnwidth}{|l|X|X|X|}
%\hline
%Content & Demand cell 1 & Demand cell 2 & Needed resources\tabularnewline
%\hline\hline
%$m_1$ & $\pi^{c_1}_{m_1}=10$ & $\pi^{c_2}_{m_1}=0$ &$\rho^{c_1}_{m_1}=\rho^{c_2}_{m_1}=6$ \tabularnewline
%\hline 
%$m_2$ & $\pi^{c_1}_{m_2}=0$ & $\pi^{c_2}_{m_2}=9$ & $\rho^{c_1}_{m_2}=\rho^{c_2}_{m_2}=6$\tabularnewline
%\hline
%$m_3$ & $\pi^{c_1}_{m_3}=6$ & $\pi^{c_2}_{m_3}=6$ & $\rho^{c_1}_{m_3}=\rho^{c_2}_{m_3}=6$\tabularnewline
%\hline
%\end{tabularx}
%\end{table}
%
%Suppose we have two neighboring cells~$c_1$ and~$c_2$ and three content
%items $m_1$, $m_2$, and $m_3$, whose demand is summarized in
%Tab.~\ref{tab:example}. We could create an area including~$c_1$ and
%broadcast content~$m_1$ therein, obtaining a satisfaction of~10 (let us
%assume there is no unicast demand). Then, we would try to create another
%area with~$c_2$ and broadcast~$m_2$ therein, but doing so would violate
%constraint~(\ref{eq:constr-R-ex}). Instead, we would be better off
%creating a single area, include both cells therein, and broadcast
%content~$m_3$, obtaining a total profit of~12.

%This simple example demonstrates the kind of tradeoff -- small areas with high interference
There is an essential tradeoff between two choices: small areas with high interference
or bigger areas with less interference but broadcasting less popular content -- we have to
deal with when forming the broadcasting areas. There are two main
approaches we can adopt to solve the problem, 
which we name {\em merge} and {\em grow}.

\subsubsection{The {\em merge} approach}

The intuition behind this approach comes directly from the above discussion. We start from
an assignment where we create an area per cell (line~\ref{line:merge-new}). Then, we merge neighboring areas so as to
maximize the (immediate) performance improvement (line~\ref{line:merge-argmax}). We stop when both the following conditions
are met: first, the number of areas is below the maximum limit $\hat{A}$ (i.e.,~256); second, there are no more
pairs of areas that can be merged increasing the performance (line~\ref{line:merge-break}). More formally, we proceed as shown in Algorithm~\ref{alg:merge}.

\begin{algorithm}
\begin{algorithmic}[1]
\Require{$\Cc$}
\State{$\Ac\gets\emptyset$}
\ForAll{$c\in\Cc$}
\State{$a\gets\{c\}$} \label{line:merge-new}
\State{$\Ac\gets\Ac\cup\{a\}$} \label{line:merge-add}
\EndFor
\While{True}
\State{$(a_1,a_2)\gets\arg\max_{\Ac^2\colon a_1\in \Nc^{a_2}}\texttt{pr\_merge}(a_1,a_2)$} \label{line:merge-argmax}
\If{$|\Ac|\leq \hat{A} \wedge\texttt{pr\_merge}(a_1,a_2) \leq 0$}
\State{\textbf{break}} \label{line:merge-break}
\EndIf
\State{$a_1\gets a_1\cup a_2$} \label{line:merge-merge}
\State{$\Ac\gets\Ac\setminus\{a_2\}$}
\EndWhile
\State\Return{$\Ac$}
\end{algorithmic}
\caption{\label{alg:merge}Merge approach}
\end{algorithm}

%More formally, we proceed as shown in Algorithm~\ref{alg:merge}. We create an area for each cell
%(line~\ref{line:merge-new}), and add such areas to~$\Ac$ (line~\ref{line:merge-add}).
%Then, we select the pair~$a_1,a_2$ of neighboring areas whose merge would maximize the profit
%(line~\ref{line:merge-argmax}). (In line 6, we slightly abuse the notation and
%denote by $\Nc^a$ the set of  neighboring areas of $a$.)
%%
%If the number of existing areas is not too big and merging~$a_1$ and~$a_2$ would bring no benefit,
%we stop the algorithm (line~\ref{line:merge-break}). Otherwise, we merge~$a_1$ and~$a_2$
%(line~\ref{line:merge-merge}).

It is important to stress that evaluating the profit~\texttt{pr\_merge} does not necessarily mean
computing the full performance function~(\ref{eq:performance}). Indeed, we can resort
to simpler proxy functions, as detailed in Section~\ref{ssec:metrics}.

\subsubsection{The {\em grow} approach}

The merge approach above is very simple; indeed, we perform a single operation -- merge
two areas -- until the termination condition is reached.
Simplicity is, in general, a good thing; however, some scenarios may call for
a higher level of flexibility. In the following, we present an alternate approach,
called {\em grow}.

%In the grow approach, each step consists of two phases. In the first phase, we identify a cell
%that is suitable to become part of a new area. In the second one, we add new cells to such a newly
%created area.

\begin{algorithm}
\begin{algorithmic}[1]
\Require{$\Cc$}
\State{$\Ac\gets\emptyset$}
\While{\textbf{True}} %$}
\State{$c^{\star}\gets\arg\max_{c\in\Cc}\texttt{pr\_create}(c)$}\label{line:grow-cstar}
\If{$|\Ac|\leq\hat{A}\wedge\texttt{pr\_create}(c^{\star})>0$} \label{line:grow-stop}
%\State{\textbf{break}}
\EndIf
\State{$a\gets\{c^{\star}\}$}\label{line:grow-makearea}
\While{$|a|\leq|\Cc|$}\label{line:grow-while}
\State{$c^{\prime}\gets\arg\max_{c\in\Cc\colon c\in N_{a}}\texttt{pr\_add}(c,a)$}\label{line:grow-cprime}
\State{\textbf{if}~$\texttt{pr\_add}(c,a)\leq 0$~\textbf{then break}}\label{line:grow-break}
\State{$a\gets a\cup\{c^{\prime}\}$}\label{line:grow-add}
\EndWhile
\State{$\Ac\gets\Ac\cup\{a\}$}
\State{\textbf{else break}}
\EndWhile
\State\Return{$\Ac$}
\end{algorithmic}
\caption{\label{alg:grow}Grow approach}
\end{algorithm}
We select the cell~$c^{\star}$ that is best suited for a new area in line~\ref{line:grow-cstar},
and create a new area containing only this cell (line~\ref{line:grow-makearea}).
Next, we grow the newly created area by selecting (line~\ref{line:grow-cprime}) a cell~$c^{\prime}$
to add. $c^{\prime}$~is the cell, among the ones neighboring with area~$a$, that is most profitable
to add. If the profit is negative, then there are no more cells we can add to~$a$
(line~\ref{line:grow-break}), and we add~$a$ to the set~$\Ac$ and move on creating the next area.

This approach is more complex than the merge one, because of the two-phase
structure of each step. However, with such a complexity comes a better flexibility, e.g.,
in defining cell-to-area assignments where areas overlap and there are cells
not included in any area.

Similar to the previous case, notice that we have not given a definition of the~\texttt{pr\_add}
and~\texttt{pr\_create} profit metrics. Different metrics can be adopted while pursuing different
trade-offs between effectiveness and efficiency, as explained next.

\subsection{Profit metrics
\label{ssec:metrics}
}

%While describing the merge and grow approaches, we have used expressions like ``the most
%profitable areas to merge'', ``the most profitable cell to add'', and so on. We used the
%functions~\texttt{pr\_merge},~\texttt{pr\_add} and~\texttt{pr\_create} in Algorithm~\ref{alg:merge}
%and Algorithm~\ref{alg:grow}, but gave no definition of what such ``profits'' are.

Profit metrics are evaluated very often during the execution of those algorithms; therefore,
it is of paramount importance that they can be computed efficiently. However, such metrics
must also represent a good proxy of the performance metric in~(\ref{eq:performance}).

There are two fundamental ways of defining profit. One is considering all such aspects as
interference and propagation, and this essentially means computing~(\ref{eq:performance}) every time.
The other is focusing on content demand, with the assumption that it is the main
factor to account for in order to maximize performance.

\subsubsection{Demand-based profit}

Content demand~$\pi^c_m$, i.e., the number of users in a cell interested
in a certain content, is obviously
the main factor to account for when taking such decision as creating or merging areas. For
the sake of simplicity, we may decide to make it the {\em sole} factor to look at.

The~\texttt{pr\_merge} function used in line~\ref{line:merge-argmax} of Algorithm~\ref{alg:merge}
is thus defined as follows:
\begin{equation}
\label{eq:demandbased-profit-merge}
\texttt{pr\_merge}(a_1,a_2)=\frac{1}{U^{a_1}+U^{a_2}}\max_{m\in\Mc}\left(\sum_{c\in a_1\cup a_2}\pi_m^c\right)
\end{equation}
where $U^{a}$ denotes the number of users in area $a$. 
Equation~(\ref{eq:demandbased-profit-merge}) says that we seek to merge those areas with a very
strong interest in the same content (as opposed to a weaker interest for different
content).

Similarly, the~\texttt{pr\_create} function used in line~\ref{line:grow-cstar} of
Algorithm~\ref{alg:grow} is:
\begin{equation}
\label{eq:demandbased-profit-create}
\texttt{pr\_create}(c)=\max_{m\in\Mc}\pi_m^c.
\end{equation}
In~(\ref{eq:demandbased-profit-create}), we simply select the most popular content in cell~$c$.
Therefore, we tend to create new areas in those cells where there is a clearly prominent
content.

Finally, the~\texttt{pr\_add} function used in line~\ref{line:grow-cprime} accounts for
how popular the content~$\mu_a$, broadcasted in area~$a$, is in cell~$c$, as shown
in~(\ref{eq:demandbased-profit-add}):
\begin{equation}
\label{eq:demandbased-profit-add}
\texttt{pr\_add}(c,a)=\pi_{\mu_a}^c.
\end{equation}

Using the definitions above implies that side effects such as
interference are neglected, but has a
clear performance advantage. Content demand and interest are known a priori, and are not
influenced by our decisions. Therefore, identifying and evaluating the possible actions
is remarkably simple -- and, thus, fast.

\begin{figure}
\centering
\includegraphics[width=.18\textwidth]{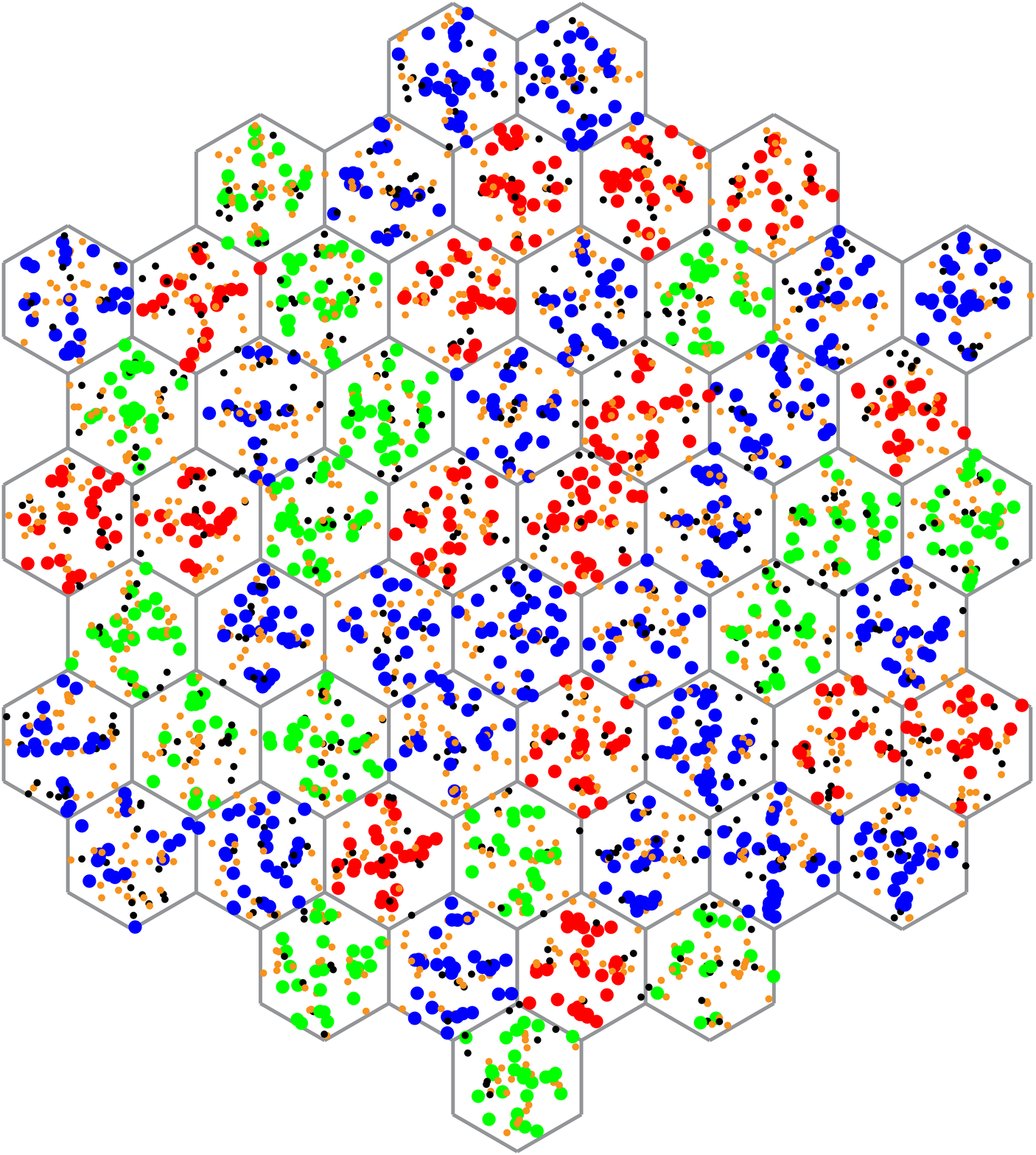}
\caption{Topology and demand: red, green and blue points correspond to {\em streaming} content,
orange dots to the {\em update} one.
\label{fig:topology}}
\end{figure}

\subsubsection{Holistic profits}

The opposite approach consists in considering all the consequences of, say, adding a cell~$c$
to an area~$a$, i.e., in accounting not only  for the popularity of
the content in the cell, but also  for how
the performance of other users, e.g., unicast ones, is affected.

This means to proceed as follows:
\begin{enumerate}
\item taking an action, e.g., adding cell~$c$ to area~$a$ or merging two areas $a_1,a_2$,
and updating the~$\alpha$-values accordingly;
\item running Algorithm~\ref{alg:decide-content} on the resulting cell-to-area assignment;
\item recomputing the global score through~(\ref{eq:performance}), as explained in 
Section~\ref{sec:model}.
\end{enumerate}
%On the one hand, taking such an approach to assess the profit
%associated with an action, does come at a significant cost in terms of
%computational complexity. 
%On the other hand, we are guaranteed not to take actions with unexpected consequences, e.g.,
%in terms of generated interference.

\section{Results
\label{sec:results}
}

Here, we first describe the network and traffic scenario that we used
to derive performance results, then we present a  comparison among the
approaches introduced 
above. 

\subsection{Reference scenario}

We evaluate our algorithms in a large-scale scenario typically used for
3GPP evaluation~\cite{scenario}. The scenario comprises a service area
of 12.34~km$^2$, covered by 57 cells deployed at 19 tri-sectorial sites.
There are a total of 3420~users, uniformly distributed under the cell
coverage areas. Content is available as either {\em update} or 
{\em streaming}. The former, available as a single item, 
could represent a local map update, 
and it is less resource demanding. The latter could
be seen as news clips streamed to users, and it is obviously more
resource-demanding; we will consider three different items of streaming type.

We focus on a single time period, and assume that each user is interested in exactly one 
broadcast content item, as follows:
\begin{itemize}
\item with 20\%~probability, the user requests the {\em update} item;
\item with 80\%~probability, the user requests one among 
{\em streaming1}, {\em streaming2} or {\em streaming3} items.
\end{itemize}
Furthermore, streaming items are location dependent, i.e., in each
cell users may be interested in only one of the three item. The {\em
  update} content, instead, 
is requested throughout the whole topology -- but with lower probability.

There are a total of~$R=500$ resources available per frame, each
corresponding to an 
RB in LTE. At most  60\% of such resources, i.e.,~$r=300$, can be
used for broadcast. The minimum amount of resources needed to broadcast
a content is~$\rho^c_m=120$ for {\em streaming} content, and~$\rho^c_m=80$ for
the {\em update} content.
Topology and demand are summarized in Fig.~\ref{fig:topology}.

\begin{figure}[t]
\centering
\subfigure[\label{fig:base-score}]{
\includegraphics[width=.22\textwidth]{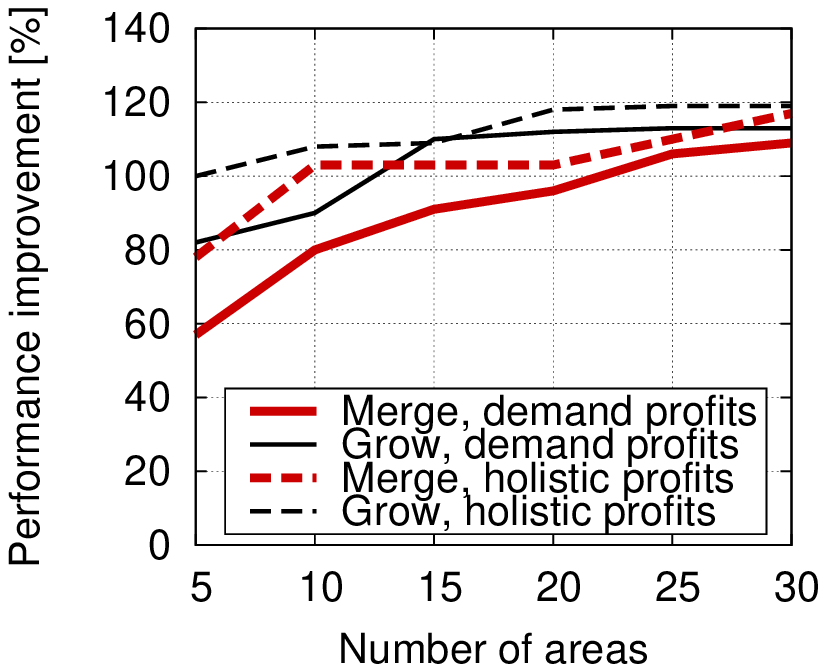}}
\subfigure[\label{fig:base-size}]{
\includegraphics[width=.22\textwidth]{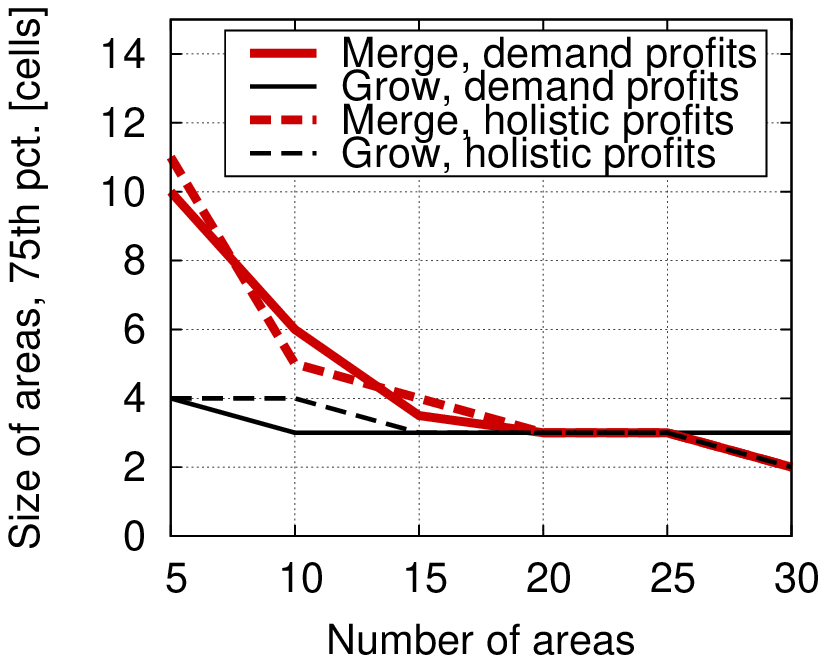}}
\caption{Performance improvement (a) and area size (b) as  functions of the number 
of areas, for different approaches and profit metrics.
}
\end{figure}

\begin{figure}[b]
\centering
\subfigure[\label{fig:base-solutions-join}]{
\includegraphics[width=.18\textwidth]{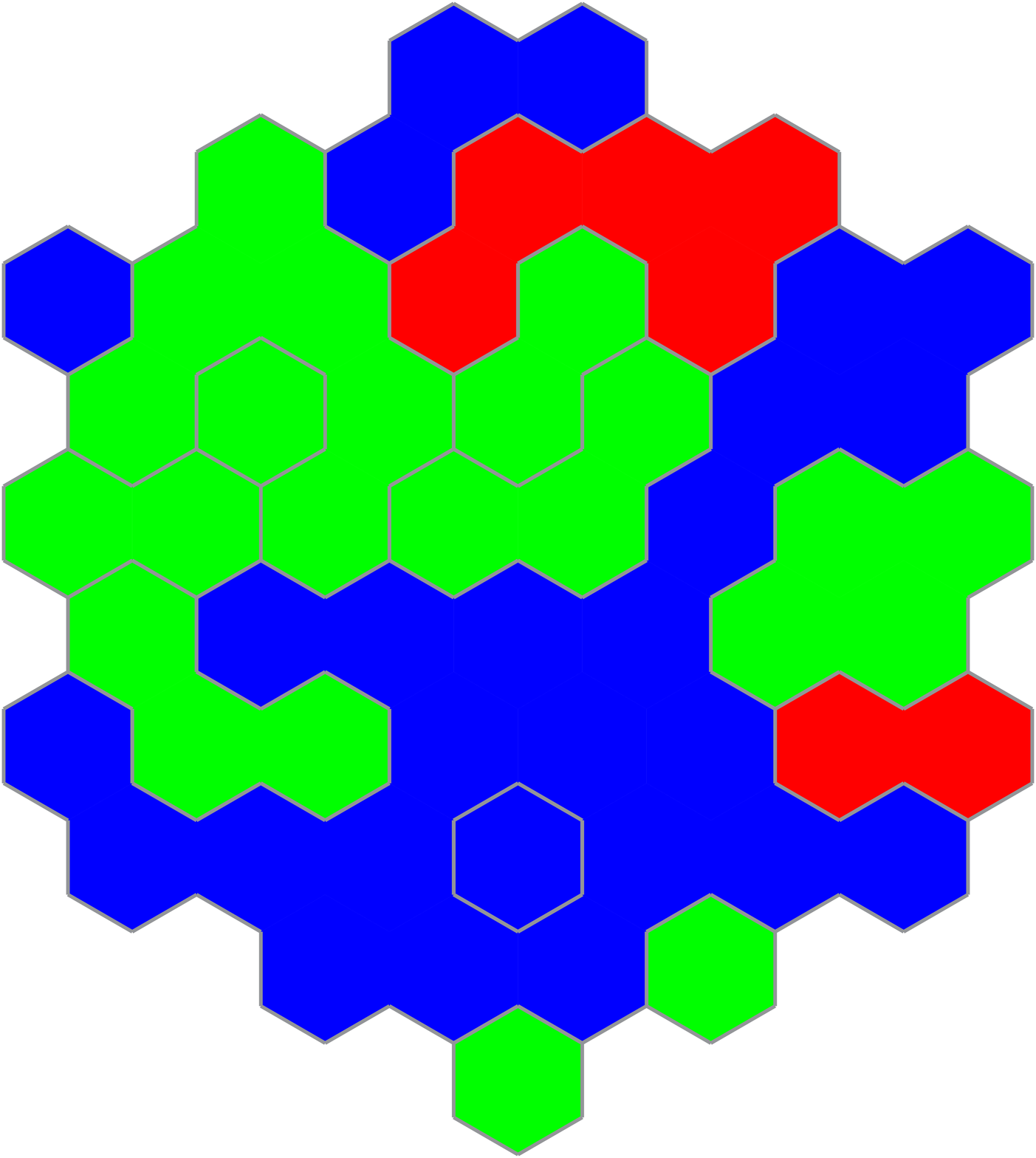}}
\subfigure[\label{fig:base-solutions-grow}]{
\includegraphics[width=.18\textwidth]{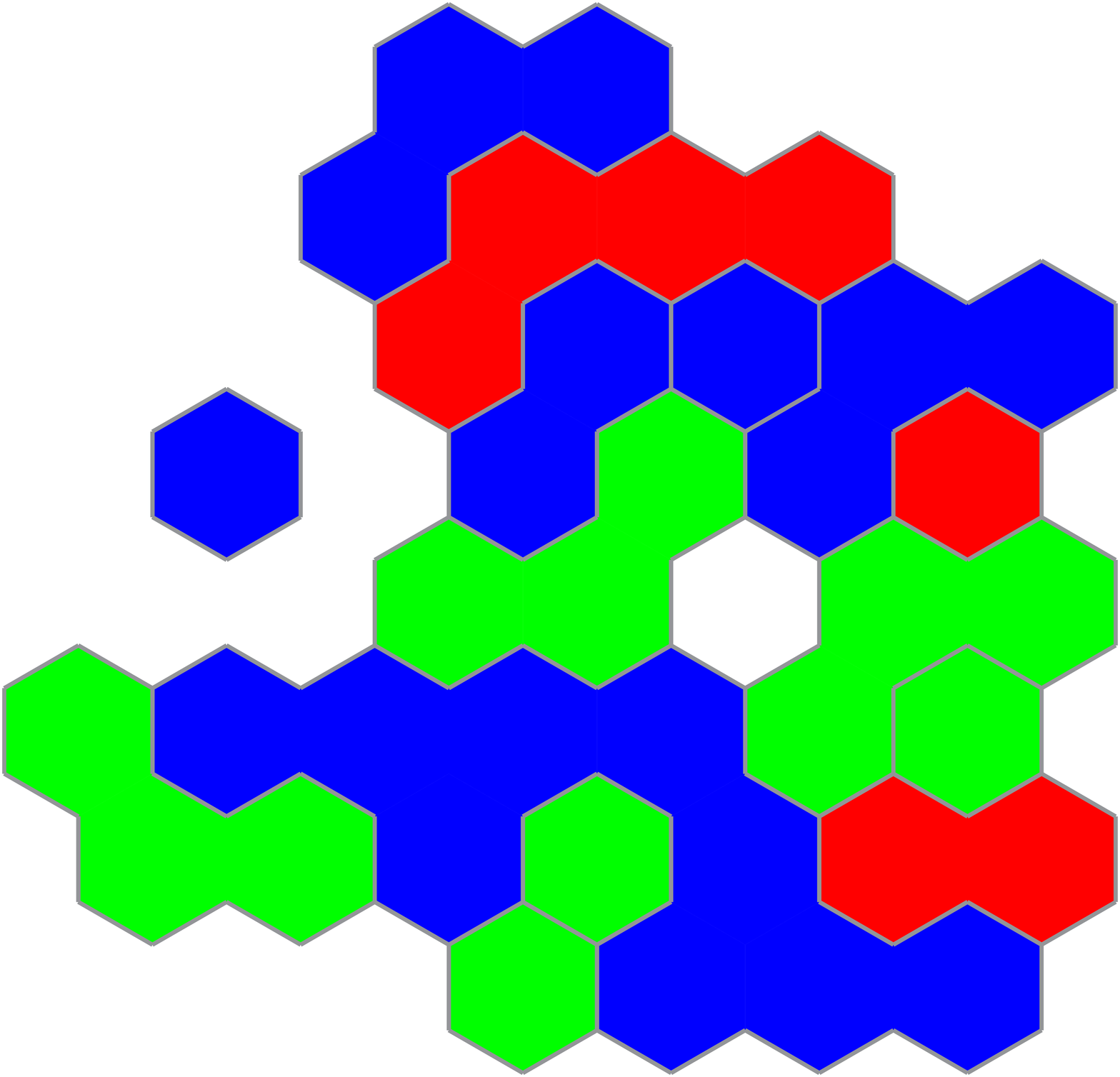}}
\caption{
Solutions yielded by the {\em merge} (a) and {\em grow} (b) approaches,
for a maximum of~10 areas.
\label{fig:base-solutions}
}
\end{figure}

\subsection{Performance of the {\em grow} and {\em merge} approaches}

The first thing we are interested in is the effectiveness of the two
approaches we described in Section~\ref{sec:algos}, i.e., {\em grow} and
{\em merge}. We take as a reference the performance, computed
through~(\ref{eq:performance}), when broadcasting is disabled,
i.e.,~$\alpha^c_a=0,\forall c,a$. Then, we measure how much such a
performance is improved by enabling broadcasting, and using either
approach to form the area. The content to broadcast in each area is
always chosen through Algorithm~\ref{alg:decide-content}.

We vary the number of allowed areas, $\hat{A}$, between~5 and~30. These numbers are
significantly smaller than the limit of~256 areas in the
standard since our topology is much smaller than typical LTE
geographical areas, which can span over hundreds of square kilometers.

Fig.~\ref{fig:base-score} shows the performance improvement (with respect to
the no-broadcast case) we obtain with the different approach and profit
metrics.

A first observation we can make concerns the influence of the number
maximum of areas: the bigger it is, the better the performance. This is
expected; intuitively, more areas entail more flexibility. 
%Looking at the
%demand in Fig.~\ref{fig:topology}, we can imagine that we could try to
%form an area for each group of contiguous cells requesting the same {\em
%streaming} content -- if possible. As we will see later, such an
%intuition is not far from reality.

Moving to the approaches, we observe that the {\em grow} approach
consistently performs better than the {\em merge} one. As expected,
enjoying a higher level of flexibility pays off. Less obviously, moving
from the demand-based profit metric to the holistic one translates into
a significant performance improvement only for the {\em merge} approach,
and only when the limit on the number of areas to create is very tight.

%This is good news. It means that, if the algorithm we employ to form the
%areas is smart enough, we do not need to resort to costly, detailed
%performance metrics. This, in turn, allows us to schedule and
%re-schedule content broadcasting with remarkable speed -- in real time
%if need be.

Now, we have to determine why the {\em grow} approach performs better.
Fig.~\ref{fig:base-size} gives us an answer: it forms smaller areas.
Recall the discussion in Section~\ref{sec:algos} about the difference
between the two approaches: with the {\em merge} approach, we are bound
to put every cell in (exactly) one area. This may not sound like a bad idea
in our scenario; after all, we have a significant demand for
broadcast-able content throughout the whole topology.

It turns out, instead, that insisting to have all cells assigned to an
area severely impairs performance. We can get an idea of why this
happens by looking at Fig.~\ref{fig:base-solutions}. As expected, the
{\em merge} approach yields a solution where all cells belong to an area
(Fig.~\ref{fig:base-solutions-join}). By comparing
Fig.~\ref{fig:base-solutions-join} to the demand in
Fig.~\ref{fig:topology}, however, we can see that many cells are in
areas that broadcast a content that nobody wants. 
%-- look, for example,
%at the cells in the large blue area at the bottom. In these cells,
%the~$\rho^c_m=120$ resources assigned to broadcasting are outright wasted.

Compare now the solution yielded by the {\em grow} approach, in
Fig.~\ref{fig:base-solutions-grow}. The first thing that strikes our
attention is that many cells do not belong to any area. Looking more
carefully, we can see that this typically happens with cells surrounded
by neighbors with different demand (look, e.g., at the ``hole'' towards
the center of the topology or the ``island'' on the left). These cells are
never selected during the {\em grow} process (Algorithm~\ref{alg:grow}), and
therefore all the resources therein are used for unicasting. As we can
see from Fig.~\ref{fig:base-score}, this is more convenient than
broadcasting content with low popularity.

\noindent
{\bf Summary.}
We can thus draw three main conclusions from our performance evaluation.
First, the {\em grow} approach outperforms the {\em merge} one, owing to
its higher level of flexibility. Second, such a flexibility is
sufficient to compensate the usage of a simpler profit metric, namely,
interest-based. Third, such a difference in performance is mostly due to
the tendency of the {\em merge} approach to assign each cell to an area,
at the cost of broadcasting uninteresting content.

\section{Conclusions
\label{sec:conclusions}
}

We have considered the broadcasting features in LTE.
%, one of the most
%promising ways to efficiently deliver real-time streaming to a large
%number of users without overloading the network. 
Specifically, we
addressed the problems of (i)~forming the areas, i.e., decide which
cell(s) belong to each area, and (ii)~deciding which content to
broadcast in each cell.
%We began by presenting a simple model that can capture all meaningful
%aspects of the (fairly complex) system model specified by 3GPP
%standards, including interference.
We
% then 
decoupled the LTE broadcasting problems of forming the areas and choosing the
content to broadcast. We presented a simple and straightforward strategy
for the last problem, and two approaches, presenting different levels of
complexity and flexibility, for the first. 
Additionally, we described two
ways of assessing the profitability of possible assignments:
accounting for content demand alone, or adding interference issues to
the picture.
By evaluating our system in a large-scale, real-world scenario, we
found that selecting the most flexible approach makes it possible to use
the simplest profit metric, thus being able to (re)schedule the content
to broadcast on a very fine time granularity. 
We also investigated the
reason for the difference in performance between the two approaches, and
found that trying to assign all cells to an area is harmful to the
overall performance.

\section*{Acknowledgment}
This paper was made possible by NPRP grant $\sharp\,5-782-2-322$ from the Qatar National Research Fund (a member of Qatar
Foundation). The statements made herein are solely the responsibility of
the authors.

\end{document}